\documentclass[a4paper,fleqn,usenatbib]{mnras}

\usepackage[T1]{fontenc}
\usepackage{ae,aecompl}

\usepackage{graphicx}	% Including figure files
\usepackage{amsmath}	% Advanced maths commands
\usepackage{amssymb}	% Extra maths symbols
\usepackage{wasysym}
\usepackage{stmaryrd}

\title[Tidal evolution of the Earth]{Scaling in global tidal dissipation of the Earth-Moon system}

\author[Maurice H.P.M. van Putten]{Maurice H.P.M. van Putten$^{1}$\thanks{E-mail: mvp@sejong.ac.kr}\\
$^{1}$Sejong University, 98 Gunja-Dong Gwangin-gu, Seoul 143-747, Korea}

\date{Accepted XXX. Received YYY; in original form ZZZ}

\pubyear{2016}

\begin{document}
\label{firstpage}
\pagerange{\pageref{firstpage}--\pageref{lastpage}}
\maketitle

% Abstract of the paper
\begin{abstract}
The Moon migrated to $r_{\leftmoon}\simeq3.8\times10^{10}$ cm over a characteristic time $r/v=10^{10}$ Gyr by tidal interaction with the Earth's oceans at a present velocity of $v=3.8$ cm yr$^{-1}$. We derive scaling of global dissipation that covers the entire history over the past 4.52 Gyr. Off-resonance tidal interactions at relatively short tidal periods in the past reveal the need for scaling {with amplitude}. The global properties of the complex spatio-temporal dynamics and dissipation in broad spectrum ocean waves is modeled by damping $\epsilon = h F/(2Q_0)$, where $h$ is the tidal wave amplitude, $F$ is the tidal frequency, and $Q_0$ is the $Q$-factor at the present time. It satisfies $Q_0\simeq 14$ for consistency of migration time and age of the Moon consistent with observations for a near-resonance state today. It shows a startingly fast eviction of the Moon from an unstable near-synchronous orbit close to the Roche limit, probably in a protolunar disk. Rapid spin down of the Earth from an intial $\sim30\%$ of break-up by the Moon favored early formation of a clement global climate. Our theory suggests moons may be similarly advantageous to potentially habitable exoplanets.
\end{abstract}


\section{Introduction}

Over the past eons, tidal interaction produced a major evolution of the Earth-Moon system, producing a relatively slowly rotating planet down from about one-third of centrifugal break-up at birth, very similar then to Jupiter's state today. For a detailed discussion on the break-up angular velocity $\Omega_b\simeq \sqrt{{GM}/{R^3}}$ of a planet of mass $M$ and radius $R$, see, e.g., \cite{dav99}.

Tidal interactions are inherently dissipative, determined by a phase-lag between tidal deformation and position of the perturber, i.e., a misalignment of tidal bulge relative to the Earth-Moon direction. For the Earth, dissipation is primarily in the ocean tidal flows, more so than viscoelastic deformation of the Earth's mantle \citep[e.g.][]{mun68,mun71,lam77,dic94,ray94,ray96,efr09,efr12,efr15}, whose seismic frequencies are relative high compared to the tidal frequency \citep{lov11,mun60,dah74} with the exception of those driven by ocean waves \citep{web07}. However, a detailed quantitative account for the overall Moon's migration time due to various nonlinear dissipation channels \citep[][]{sto48,sto57,mun68,egb01}, remains to be identified.

Tidal dissipation \citep[][]{mun71,web82,dic94} has various mechanisms in shallow water wave theory, some of which have recently been highlighted in detailed numerical simulations on ocean dynamics and dissipation covering relatively short initial and present epochs \citep[e.g][]{tou94,tou98,egb04,sta14}. Dissipation generally occurs when nonlinear steepening exceeds the mitigating effect of dispersion. Determining the net global result from detailed modeling of tidal dissipation is particularly challenging by the diversity of oceans and coastal regions, where most of the dissipation is expected to occur \citep{mil66,mun71}.

Here, we focus on scaling in both amplitude and frequency of global tidal dissipation to account for the Earth-Moon history over the past 4.52 Gyr. {This approach aims at providing an effective description of an otherwise complex spatio-temperal distribution of dissipation in broadband ocean waves.} For a confrontation with data, the Moon's migration time is computed by numerical integration of angular momentum transfer backwards in time, to the instant of its formation from the Earth or a surrounding proto-lunar disk. This approach enables taking into account variations in tidal implitude over a few orders of magnitude, the effect of which seems not to have been computed before. The evolution by coupling to the Earth's spin \citep[e.g.][]{efr09} is conveniently described by the orbital angular momentum $m\sqrt{GM(1-e^2)a}$ with semi-major axis $a$, where $m=7.35\times 10^{22}$ g and $M=5.97\times 10^{27}$ g denote the mass of the Moon and, respectively, Earth. The orbital ellipticity $e$ is presently about $5.5\%$. A large dynamic range in tidal interaction strength arises from the tidal amplitude $h\propto r^{-3}$ by which the Moon's specific angular momentum {at radius $r\simeq a$} evolves according to
\begin{eqnarray}
\frac{dj}{dt} \propto \frac{h}{r^3} \sin(2\Delta \varphi),
\label{EQN_j}
\end{eqnarray}
where $\Delta\varphi$ is the phase-lag of the Moon's orbit relative {to the tide raised on the Earth}, and $r^{-3}$ is the mutual interaction strength for a given $h$. As a result, $dj/dt\propto r^{-6}$ \citep[cf.][]{efr09}.

To begin, we first recall some general conditions for dissipation in ocean tidal flows (\S2). 
In \S3, we formulate our scaling of dissipation in tidal amplitude and frequency. It serves to parameterize damping in our model based on (\ref{EQN_j}) and the pendulum equation (\S4). This model is explored numerically in \S5. In \S6, we summarize the results.

\section{Some conditions for tidal dissipation}

While the theory of linear shallow water waves is dissipationless, finite amplitude waves can steepen to dissipative bores, provided that steeping exceeds the mitigating effect of dispersion. {The degree of nonlinearity over dispersion is expressed by Ursell number $Ur=h\lambda^2/d^3$, by the amplitude $h$ and wave length $\lambda$ \citep{urs53,mun68,bar04}}
\begin{eqnarray}
\mbox{Ur} = \frac{h g P^2}{ d^2} \simeq 300 \times \left(\frac{h}{0.3\,\mbox{m}}\right)\left(\frac{d}{4000\,\mbox{m}}\right)^{-2}
\left(\frac{P}{12\,\mbox{hr}}\right)^2 > 1,
\label{EQN_Ursell}
\end{eqnarray} 
here specialized to shallow water waves with propagation speed $c=\sqrt{gd}$ in oceans of depth $d$ at tidal period $P$, where $g=9.8$\,m\,s$^{-2}$ is the Earth's gravitational acceleration. {While, a large Ursell number is suggestive of a general tendency for wave breaking, it does not directly define scaling of dissipation.}

For breaking to occur, {steepening} must be sufficiently fast \cite[e.g.][]{mun68} (see \cite{lev77} for a discussion in Burgers' equation with time-dependent forcing). In nonlinear wave motion \citep[cf.][]{whi74}, it results from steepening after a time
\begin{eqnarray}
\frac{t_*}{P}\simeq -\left(\frac{\partial c}{\partial x}\right)^{-1} \simeq \frac{c^2}{gh}=\frac{d}{h}.
\label{EQN_wb}
\end{eqnarray}
(For a more detailed discussion, see \cite{sto57}.) {Here, (\ref{EQN_wb}) is relaxed by an additional $Q$-factor of the oceans with reflection of tidal waves off coastal regions. With $d/h=O(10^4)$, however, wave breaking is unlikely to occur in the open oceans even at present-day values of $Q$ (\S3 below).} Instead, it believed to occur in shallow seas, in run-up waves in shoaling shelf regions with slopes $s$ satisfying \citep{mun68,bar04,sal16} 
\begin{eqnarray}
\chi \simeq 20 \times  \left(\frac{s}{0.01}\right) \left(\frac{h}{0.3\,\mbox{m}}\right)^{-\frac{1}{2}}\left(\frac{P}{12\,\mbox{hr}}\right)^\frac{1}{2}\left(\frac{d}{4000\,\mbox{m}}\right)^\frac{1}{4} < 1.9.
\label{EQN_surf}
\end{eqnarray}

The dissipation rate of bores produced by wave breaking satisfies entropy creation in shocks of compressible gas dynamics, i.e., scaling with the cube of their amplitude. By aforementioned $h\propto r^{-3}$, wave breaking, by either (\ref{EQN_Ursell}) or (\ref{EQN_surf}), provides a time rate of dissipation effectively described by damping proportional to $h$, possibly including tidal frequency in light of broadband wave spectra. Our approach is focused on scaling in global dissipation described by (\ref{EQN_epsD0}), to capture the net result of an otherwise complex spatio-temporal distribution of dissipation in the Earth's oceans.

\section{Amplitude-frequency scaling of dissipation}

{We set out model global tidal dissipation in the Earth's oceans by (\ref{EQN_j}) in dimensionless variables, normalized by today's Moon migration data and age. The tidal amplitude $h$ is described by a damped linear pendulum equation with eigenfrequency $\omega_0$, forced at semi-diurnal tidal frequency $\omega^\prime$ by the action of the Moon at orbital angular velocity $\omega$,
\begin{eqnarray}
\omega' = 2(\Omega - \omega),
\label{EQN_ot}
\end{eqnarray}
where $\Omega$ denotes the angular velocity of the Earth.
The damping coefficient $\epsilon$ in the pendulum equation represents dissipation with associated phase lag $\Delta\varphi$ and $Q$-factor $Q=1/(2\epsilon)$.}

To study the net result over the entire history of the Earth-Moon system, we consider the general scaling in amplitude $h$ and dimensionless tidal frequency $F$,
\begin{eqnarray}
\epsilon \propto \epsilon_0 h^pF^q
\label{EQN_epsD0}
\end{eqnarray}
and its confrontation with data for various choices of $p,q=0,1$. 

{With (\ref{EQN_epsD0}), we aim to capture the net result of tidal amplitude-frequency dependence that covers weak and strong interactions at present and, respectively, back in the distant past.} In the Stokes' limit $p=q=0$, for instance, and ignoring resonances, integration of (\ref{EQN_j}) backwards in time obtains a migration time on the order of 1 Gyr at odds with the Moon's age \citep[cf.][]{ger55,gol66,mun68}. This suggests that tidal dissipation is anomalously high at present. (Equivalently, the $Q$-factor was higher in the past, e.g., \cite[][]{tou98}.) Taking into account the inertial range off-resonance in the past (Fig. 1), however, one is led to the opposite conclusion with a previously {\em low} $Q$-factor, here revisited by numerical integration of (\ref{EQN_j}) with (\ref{EQN_epsD0}) in the forced pendulum equation.

The Moon migrated to its present mean distance $r_{\leftmoon}=3.8\times 10^{10}$ cm with a radial velocity $v=3.8$ cm yr$^{-1}$ \citep{bil99}. It defines a characteristic migration time scale
\begin{eqnarray}
t_m = \frac{r_{\leftmoon}}{v} \simeq 10 \,\mbox{Gyr},
\label{EQN_j0}
\end{eqnarray}
that will serve to express the equation of motion (\ref{EQN_j}) in terms of a dimensionless time $t/t_m$ (\S3 below). Being tidally locked, the Moon spins slowly with negligible angular momentum. By now, the Earth's spin has effectively been transferred to the Earth-Moon orbit, since
\begin{eqnarray}
\frac{I\Omega}{mj}\simeq0.8247\left(\frac{M}{m}\right)\left(\frac{R}{r}\right)^2\left(\frac{\Omega}{\omega}\right)\simeq 0.2062
\label{EQN_r4}
\end{eqnarray} 
based on the Moon's mass $m\simeq M/81$, the Earth's angular momentum $I\Omega$, radius $R_\varoplus=6\times 10^8$ and moment of inertia
$I=0.8247 I_0\left[1+0.6380\left({\Omega}/{\Omega_b}\right)^2\right], $
$I_0=(2/5)MR_\varoplus^2$ \citep{gol66,rom77}. 

The present-day tidal deformation $h_0$ (in cm) has a characteristic scale defined by the zero-frequency perturbation of the Newtonian binding energy $U_N$ of self-gravity and $U_t$ in the Moon's tidal field,
$\delta U_b = {GM^2}{R^{-3}_\varoplus} h_0^2$, $\delta U_t = {GmMR_\varoplus}{r^{-3}}h_0,$
which recovers the familiar scaling
\begin{eqnarray}
h_0 = \left(\frac{R_\varoplus}{r}\right)^3 \left(\frac{m}{M}\right) R_\varoplus \simeq 30\,\mbox{cm}
\label{EQN_h0}
\end{eqnarray}
with $\delta U_b=\delta U_t\simeq 8\times 10^{24}$ erg, consistent with existing estimates on ocean tide energies \citep{mun71} and measured mean values of the dynamical tidal amplitude $h$ to about 20\% \citep{mil66,wah95}. Consistency of $h_0$ with dynamical tidal amplitude is somewhat coincidental in view of the relatively minor deformation of the Earth's mantle \citep{wah95} and an appreciable $Q$-factor in the ocean tides \citep[e.g.][]{web82}.

\begin{table*}
	\centering
	\caption{List of selected symbols used in scaling of dissipation at high Ursell numbers with tidal amplitude and frequency covering the complete evolution of the Earth-Moon system.}
	\label{tableB}
	\begin{tabular}{lllll}
		Quantity 	& symbol & normalized expression & present value(s) & comment  \\
		\hline\hline
		Earth mass                                   &  $M$                    &                    & $5.97\times 10^{27}$ g\\
		Earth radius                                 &  $R_\varoplus$     &                    & $6.37\times 10^{8}$ cm\\
		Moon mass                                  &   $m$                   &   $m/M$       & 1/81 \\
		Moon distance                             &   $r$                    &                     & $r_{\leftmoon}=3.8\times 10^{10}$ cm\\
		Ursell number                              & Ur                        & $hgP^2d^{-2}$&$10^2 < \mbox{Ur} < 10^{10}$& (\ref{EQN_Ursell})\\	    
		Earth angular velocity                  &  $\Omega$           & $\Omega/\Omega_b$ & 6\% \\
		Moon orbital angular velicity        & $\omega$            &    $\omega/\Omega$       & 1/30\\	
		Look back time                             & $\tau$                 & $(t_0-t)/t_m$                   & 0 & (12)\\
		Moon specific angular momentum &  $j$       & $j_0\le j \le j_1$  & 1  & $F(j_i)=0$ ($i=0,1)$\\       		
		tidal angular velocity                     &  $\omega^\prime$      & $0\le F(j) < 3$  & 1  & (\ref{EQN_ot}-\ref{EQN_Fj}), Fig. \ref{figFj}\\
		proximity to resonance today        &  $k_0$ &   $\omega^\prime_0/{\omega_0}$  & $0.95 < k_0 < 1.35$ & (\ref{EQN_k0}), Figs. 1, 5 \\   		
		ocean depth                                  &  $d$ & & $\sim 4$ km & (\ref{EQN_wb}), (\ref{EQN_ave})\\
		ocean $Q$-factor                         &  $Q$ & $2\pi \delta U_bD^{-1}P^{-1}$ & $5 < Q_0 < 30$ & (\ref{EQN_Q}), (\ref{EQN_QE}), Fig. 5\\
		damping coefficient                      &     $\epsilon$   & $\epsilon_0 h^p F^q$ ($p,q=0,1$)       &    $\epsilon_0 <<1$   &	(\ref{EQN_epsD0}), (\ref{EQN_epsB}-\ref{EQN_epsD}),(\ref{EQN_epsF}), Table II \\
		\hline
	\end{tabular}
\end{table*}

By the first law of thermodynamics, the Earth's rotational energy $E_{rot}=\frac{1}{2}I\Omega^2$ is deposited into the Moon's orbit, $H\simeq -GMm/(2a)$ and heat, the latter by tidal dissipation in the Earth
\begin{eqnarray}
D = \left(\Omega - \omega\right) \frac{dJ_{\leftmoon}}{dt} \simeq 3\,\mbox{TW},
\label{EQN_D}
\end{eqnarray}
i.e., about 6.6 mW\,m$^{-2}$, where 1TW=$10^{12}$ Watt inferred from the time rate-of-change in the angular momentum $J_{\leftmoon}=mj$ of the Moon \citep[][]{mun71,web82,dic94}. Presently, $\Omega/\omega\simeq 30$, whereby $E_{rot}$ is mostly dissipated into heat, powering $H$ at merely three percent efficiency. In the Earth's tides, estimates of associated $Q$-factor \citep[e.g.][]{mac64,gol66a,mun68,gol66,mun71,tyl08} satisfy
\begin{eqnarray}
Q_0 = \frac{2\pi \delta U_b}{DP} \simeq {\cal O}\left(10^1\right)
\label{EQN_Q}
\end{eqnarray}
for the present-day semidiurnal tidal period $P\simeq 12$ hr. 

\begin{figure}
%	\centerline{\includegraphics[scale=0.22]{m_moon_C}}
	\centerline{\includegraphics[scale=0.44]{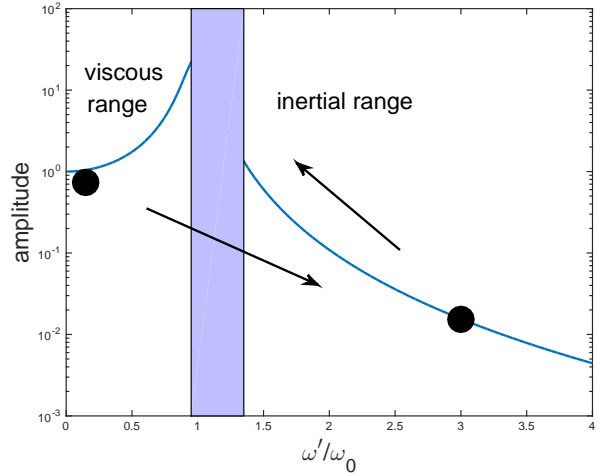}}
	\caption{Trajectory of the Earth-Moon system in the amplitude-frequency diagram of a forced pendulum with resonance frequency $\omega_0$ of today's oceans. Starting from an unstable synchronous orbit with vanishing tidal angular frequency $\omega^\prime$, the trajectory reached a state of maximal tidal frequency in the inertial range $\omega^\prime/\omega_0>>1$. As the Earth continued to spin down, the trajectory approached near-resonance today.
	{The vertical rectangle indicates the domain of present state $\omega^\prime_0/\omega_0$, that can be reached by integrating (\ref{EQN_r4e}) over a period of 4.52 Gyr.}}
	\label{figA}
\end{figure}

Our starting point to model tidal dissipation driving (\ref{EQN_j}) is the theory of forced damped pendulum equation with damping (\ref{EQN_epsD0}). At present, tidal forcing appears to be near its resonance frequency $\omega_0$ with tidal ocean flows effectively described by the theory of shallow water waves \citep{pek69,ger55,gol66,han82,web82,pla81}. While this gives rise to the presently strong Earth-Moon coupling close to resonance, this tidal interaction was effectively off-resonance {in most of the past \citep{web82,han82} when tidal frequencies were higher. Thus, the Earth-Moon tidal interaction has been rolling-up} towards resonance in the inertial range with forcing gradually approaching the ocean tidal eigenfrequency, {shown in Fig. \ref{figA} as a trajectory of amplitude $h$ versus $\omega^\prime/\omega_0$. Fig. \ref{figA} further shows a window of values $\omega^\prime/\omega_0$ today, permitted by integration over the past 4.52 Gyr with scalings (\ref{EQN_epsD0}), shown in Section 4 below.}

{The damping coefficient can also be expressed in terms of the decay time $\tau_d=1/\epsilon\omega_0$.} For instance, $\epsilon=0.07$, corresponds to a decay time of about 30 hours \citep{hen70,web82} with corresponding present-day $Q_0\simeq 7$ \citep[cf.][]{mac64,mun71,web73}. With tidal forcing mostly off-resonance in the past, in the inertial range of the pendulum equation \citep[cf.][]{web82}, consistency of migration time and age of the Moon obtains with an average damping coefficient considerably {\em larger} than today's value \citep[cf.][]{mun68}, i.e., $Q$ and $\tau_d$ were smaller in the past. 

{Table \ref{tableB} lists the most pertinent variables used in our analysis. Dimensionless  variables assume the value 1 at present by normalization to today's values. The exception is the 
look back time 
\begin{eqnarray}
	\tau=\frac{t_0-t}{t_m},
\label{EQN_tau}
\end{eqnarray}
where $t_0$ denotes the present time, for use in numerical integration with the property that $\tau=0$ now and $\tau=0.452$ at the time of birth of the Moon.}

\section{Forced damped pendulum equation}

The tidal excitation $h(t)$ (cm) can be described by the damped linear pendulum equation 
\begin{eqnarray}
\ddot{h}(t) + 2i\epsilon\omega_0 \dot{h}(t) + \omega_0^2h(t) = \omega_0^2 A \sin(\omega' t),
\label{EQN_PE}
\end{eqnarray}
where $A$ refers to the tidal interaction strength and $h_0$ in (\ref{EQN_h0}) provides a scale for $h(t)$ today, but not in the distant past. Forcing of the quadrupole tidal deformation has strength $A\propto r^{-3}$ and frequency (\ref{EQN_ot}).
Shown in Fig. \ref{figA}, (\ref{EQN_PE}) features distinct limits to the left or right of its eigenfrequency $\omega_0$ of phase-lag $\Delta\varphi$ (and hence dissipation), proportional to $\epsilon(\omega/\omega_0)$ and $\epsilon(\omega_0/\omega)$ at low, respectively, high frequency. The amplitude approaches a constant or decays with $(\omega_0/\omega)^2$ in the same two limits. 

For a dimensionless equation of motion, we define the normalized tidal frequency (\ref{EQN_ot})
\begin{eqnarray}
F(j)=5.016\left( 1 - 0.7938 j - 0.006833 j^{-3} \right)
\label{EQN_Fj}
\end{eqnarray}
with $F(j)=1$ at present $j=1$ conform (\ref{EQN_r4}). Thus, $F(j)$ and $j$ represent (\ref{EQN_ot}) upon dividing by their respective present-day values. Table \ref{tableB} lists some of the variables and their normalizations. For the purpose of our discussion and in light of $e<<1$ today, we here specialize to circular orbits $(r\equiv a$), ignoring a possibly large ellipticity at the initial stage from early resonances \citep{tou98,fro10}. 

\begin{figure}
	%\centerline{\includegraphics[scale=0.22]{m_moonA}}
    \centerline{\includegraphics[scale=0.44]{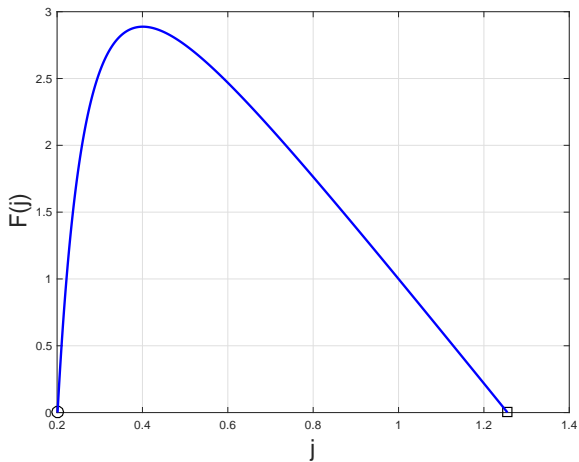}}
	\caption{Shown is the dimensionless tidal angular frequency $0\le F(j) < 3$ as a function of the dimensionless specific angular momentum $j$, satisfying $j=1$ today. It features two stationary points at the roots $F(j)=0$, corresponding to synchronous orbits $\Omega=\omega$.}
	\label{figFj}
\end{figure}

{As a function of the look back time (\ref{EQN_tau}), the evolution (\ref{EQN_j}) in dimensionless specific angular momentum $j$ satisfies the initial value problem}
\begin{eqnarray}
\frac{dj}{d\tau} = - \frac{p}{Zj^{12}}~~(\tau\ge0),~~j(0)=1,
\label{EQN_r4e}
\end{eqnarray}
where $p=\sin(2\Delta\varphi)/\sin(2\Delta\varphi_0)$ and $Z=z/z_0$ with $\Delta\varphi=\tan^{-1}(V/\left|U\right|)$ and $z=\sqrt{U^2+V^2}$, $U=F^2-k^{-2}_0$, $V=2\epsilon k^{-1}_0 F$, normalized by the present values $\Delta\varphi_0$ and $z_0$ at $\tau=0$ ($j=1$). Here, 
\begin{eqnarray}
k_0 = \frac{\omega^\prime_0}{\omega_0}
\label{EQN_k0}
\end{eqnarray} 
denotes the ratio of tidal frequency to ocean eigenfrequency today.

Evolution (\ref{EQN_r4e}) has stationary points at the two zeros of $F(j)$,
\begin{eqnarray}
j_0 = 0.2011,~j_1\simeq 1.2554
\label{EQN_jz}
\end{eqnarray}
in the past and, respectively, future when $\Omega=\omega$ (synchronous orbits, $F(j)=0$) with vanishing leading order semidiurnal tidal interaction. They are, respectively, unstable at $2.56R_\varoplus$, essentially the Roche lobe distance to the Earth, and stable at about $100R_\varoplus$. Our model, therefore, describes migration over a finite distance $r$. Normalized to $r_{\leftmoon}$ at present, it covers
\begin{eqnarray}
j_0^2 < \frac{r}{r_{\leftmoon}} < j_1^2.
\label{EQN_r12}
\end{eqnarray}
Our focus is on Eqn. (\ref{EQN_r4e}) integrated backwards in time, from $j=1$ at present down to $j_0$, at the onset of migration. Future development towards the synchronous state $j_1$ fall outside the scope of this work; cf. the synchronous Pluto-Charon system. \citep{far79,tay11,che14,tay14}.

To complete (\ref{EQN_PE}), we next turn to $\epsilon$ and consistency of
migration time and the Moon's age, estimated to be $4.53\pm0.01$ Gyr \citep{kle05} and $4.48\pm0.02$ Gyr \citep{hal08}.

\section{Numerical results}
\begin{figure*}
%	\centerline{\includegraphics[scale=0.3]{m_moonD}}
    \centerline{\includegraphics[scale=0.6]{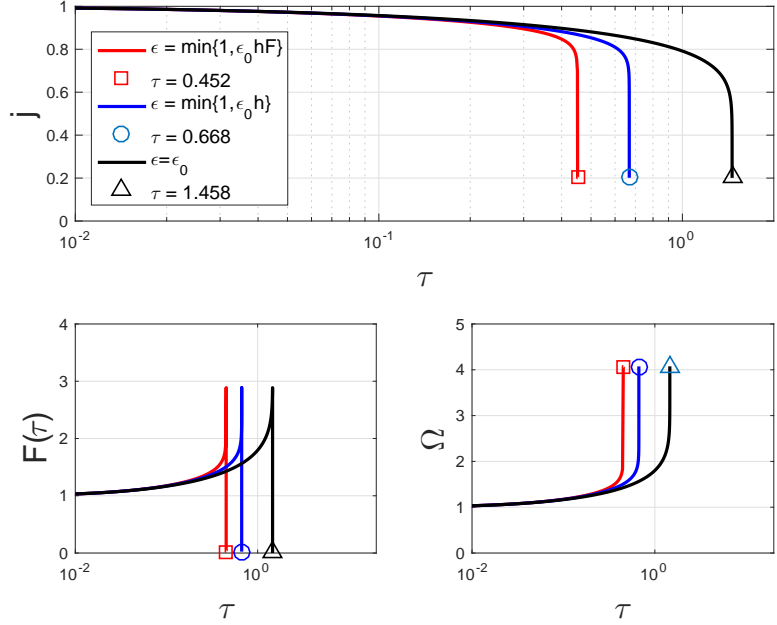}}
	\caption{Numerical integration of (\ref{EQN_r4e}) on the Moon migration with damping scaled by tidal amplitude-frequency (\ref{EQN_epsD}), tidal amplitude (\ref{EQN_epsC}) and constant damping for $\epsilon_0=0.0366$ near-resonance at $j=1$. Results are a function of look back time $\tau$ ($t=\tau t_m$) from $j=1$ at $\tau=0$ up to the first critical point $j_0\simeq 0.2011$ close to the Earth's Roche lobe distance.}
	\label{fig3}
\end{figure*}

\begin{figure*}
%	\centerline{\includegraphics[scale=0.3]{m_moonE}}
	\centerline{\includegraphics[scale=0.6]{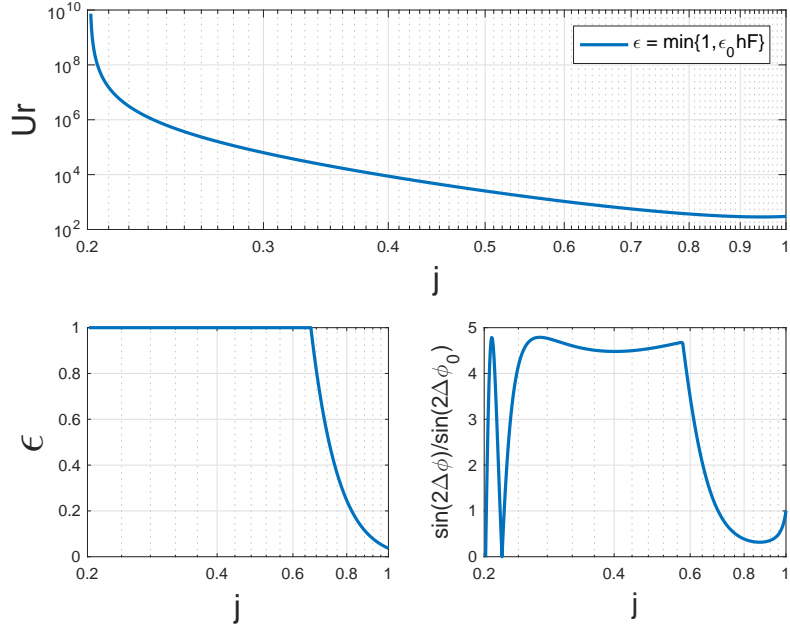}}
	\caption{The Ursell number was larger in the past, except for a 5\% dip from its current value when the Moon as 5\% closer. Nonlinear dissipation by wave breaking appears important for all of the Earth's tidal history with some suppression in the inertial range $\omega^\prime >> \omega_0$ also due to a reduced phase angle $\Delta \varphi$. Close to $j_0$, our model breaks down by essentially divergent Ursell numbers.}
	\label{fig4}
\end{figure*}

The $Q$-factor (\ref{EQN_Q}), $Q_0=1/2\epsilon_0$, of the present-day Earth-Moon tidal interaction is observationally constrained to \citep[cf.][]{hen70,web82,mac64,mun71,web73}
\begin{eqnarray}
5 < Q_0 < 30,~~0.017 < \epsilon_0 < 0.100.
\label{EQN_QE}
\end{eqnarray}
We now integrate (\ref{EQN_r4e}) over 4.52 Gyr $(\tau=0.452$) to reach a fiducial state $\omega^\prime_0/\omega_0=1.11$ of near-resonance at present. The required dissipation is compared with (\ref{EQN_QE}).

Linear theory described by constant $\epsilon=\epsilon_0$ requires $\epsilon=0.1365$. Its anomalously low $Q_0=3.7$ points to the need for enhanced damping in the past. As pointed to by the large Ursell numbers (\ref{EQN_Ursell}), we consider entropy creation in bores formed by steepening scaled by the cube of tidal amplitude, i.e., damping with tidal amplitude scaling
\begin{eqnarray}
\epsilon = \min\left\{1, \epsilon_0h \right\}.
\label{EQN_epsB}
\end{eqnarray}
A maximum of 1 is imposed for large amplitude waves, losing essentially all their energy in one cycle. Integration of (\ref{EQN_r4e}) with (\ref{EQN_epsB}) recovers a migration time equal to the Moon's age with
\begin{eqnarray}
Q_0=9.5,~~\epsilon_0=0.0527.
\label{EQN_epsC}
\end{eqnarray}
The transition to $\epsilon_A=1$ in (\ref{EQN_epsB}) is reached at 38\% of $r_{\leftmoon}$. It should be mentioned that the results are extremely insensitive to the choice of maximum in (\ref{EQN_epsB}). For instance, a maximum 2 gives $\epsilon_0=0.0527368$ instead of $\epsilon=0.0527411$ in (\ref{EQN_epsC}). While (\ref{EQN_epsC}) appears reasonable, the true $Q$-factor is conceivably larger.

Scaling characteristic of dissipation in broad spectrum wave motion further includes $F$, i.e., damping with tidal amplitude-frequency scaling
\begin{eqnarray}
\epsilon = \min\left\{1, \epsilon_0 hF \right\},
\label{EQN_epsD}
\end{eqnarray}
representing increased wave breaking overall or enhancement in high amplitude waves with $1/P$. This may derive from breaking over relatively short distances, e.g., before waves run ashore, or by flattening of ocean wave energy spectra. Without detailed numerical simulations, we resort to (\ref{EQN_r4e}) for a confrontation with data. Intregration obtains a migration time of 4.52 Gyr with (Fig. \ref{fig3})
\begin{eqnarray}
Q_0=13.7,~~\epsilon_0 = 0.0366
\label{EQN_QER}
\end{eqnarray}
consistent with (\ref{EQN_QE}). In this process, considerable suppression obtains by a reduction in phase $\Delta \varphi$ by a factor of about 3 at 75\% of $r_{\leftmoon}$ (Fig. \ref{fig4}).

Alternative scaling with tidal frequency alone,
\begin{eqnarray}
\epsilon=\epsilon_0 F
\label{EQN_epsF}
\end{eqnarray} 
as in modeling turbulent viscosity, is substantially less effective than (\ref{EQN_epsB}), requiring $\epsilon_0 = 0.0860$ ($Q_0=5.8$) for a migration time of 4.52 Gyr. For (\ref{EQN_epsD}), furthermore, Fig. \ref{fig4} confirms that the Ursell number (\ref{EQN_Ursell}) was mostly larger in the past, except for a 5\% drop at 95\% of the current Moon distance. 

\begin{figure*}
	%\centerline{\includegraphics[scale=0.3]{pars_all_b}}
	\centerline{\includegraphics[scale=0.6]{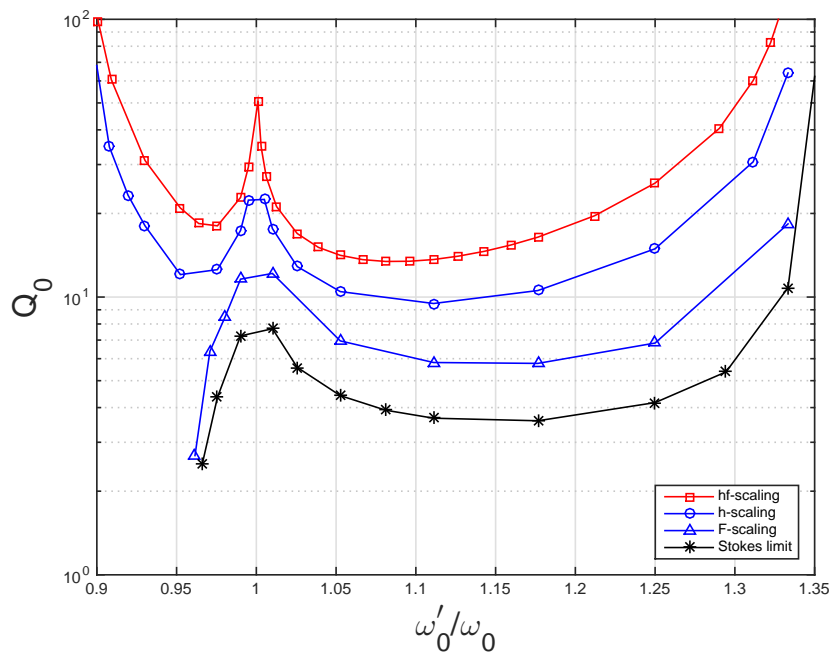}}
	\caption{Present-day $Q$-factor $Q_0$ in models of scaling with respect to tidal amplitude and frequency consistent with the Moon's age of $4.52$\,Gyr for various near-resonance conditions today, defined by tidal frequency $\omega^\prime_0$ over ocean eigenfrequency $\omega_0$ at the present time. Solutions exist for a range
	$0.95<\omega^\prime_0/\omega_0<1.35$, that defines the vertical rectangle in Fig. \ref{figA}. Values $13.4<Q_0<20$ attain for $1.01<\omega^\prime_0/\omega_0<1.25$ in evolution with $\epsilon$ scaled by both amplitude and frequency. }
	\label{figKE}
\end{figure*}

Table \ref{tableA} summarizes these numerical observations. Fig. \ref{figKE} shows the results across the full range of near-resonance states $\omega^\prime_0/\omega_0$ today as permitted by our model.

\begin{table}
	\centering
	\caption{Present-day damping coefficient $\epsilon_0$ and $Q$-factor $Q_0$ in models of scaling with respect to tidal amplitude and frequency for consistency with the Moon's age of $4.52$\,Gyr with near-resonance today ($\omega^\prime_0/\omega_0=1.11$).}
	\label{tableA}
	\begin{tabular}{lllll}
		scaling & $\epsilon$ 											& $\epsilon_0$ & $Q_0$  & Eqn.  \\
		\hline\hline
		Stokes' limit & $\epsilon_0$									   & 0.1365 & 3.7  & (\ref{EQN_PE})  \\
		tidal frequency & $\epsilon_0 F$		                              & 0.0860 &  5.8 & (\ref{EQN_epsF}) \\
    	tidal amplitude & $\min\left\{1, \epsilon_0 h \right\}$  & 0.0527 & 9.5  & (\ref{EQN_epsC}) \\
		tidal amplitude-frequency& $\min\left\{1, \epsilon_0 hF \right\}$ & 0.0366 & 13.7  & (\ref{EQN_epsD}) \\
		\hline
	\end{tabular}
\end{table}

The trajectory of (\ref{EQN_r4e}), illustrated in Fig. \ref{figA}, with amplitude-frequency scaling (\ref{EQN_epsD0}) with $p=q=1$ shows that most of the Moon's migration time is associated with the last 40\% of its current distance. The first 60\% is marked by a transition starting with a sharp spike, denoting rapid eviction of the Moon from $j=j_0$. It results from an anomalously large tidal wave amplitude of a few km (essentially the current ocean depths), defined by $h\simeq 1/j_0^{12}$ in units of the present-day amplitue scale $h_0$. Thus, $dj/d\tau = {\cal O}\left(10^8\right)$, giving an eviction time scale of about one hundred years out to tens of $R_\varoplus$. While this is unlikely accurate, e.g., tidal heating may have prevented water to condensate and may, instead, derive from a magma ocean \citep{zah15}, the idea seems valid that, once oceans form, eviction is swift.

\section{Conclusion}

The migration time of the Moon is determined by the rate of tidal dissipation, predominantly in tidal waves. In the past, forcing was mostly off-resonance in the inertial range above the resonance frequency $\omega_0$ in (\ref{EQN_PE}). {Linear theory of dissipation described by the Stokes' limit $\epsilon=\epsilon_0$} falls short of explaining a migration time equal to the Moon's age. {Nonlinear dissipation mechanisms pointed to by (\ref{EQN_Ursell}) are important}, now and even more so during off-resonance in the past. 

Dissipation in bores is similar to that in shocks of compressible gas dynamics. The same is expected to feature spectral broadening and hardening by dispersion \citep[e.g.][]{sea85,bar04}. Damping representing total tidal dissipation hereby is expected to scale both with tidal amplitude and frequency proposed in (\ref{EQN_epsD}). 

Total dissipation in the ocean tides contains also a component of internal dissipation by flows over non-smooth surfaces \citep{egb00,egb01}. It scales effectively with the cube of horizontal tidal flows \citep{gem15}, akin to high Reynolds number flows past solid objects. According to the theory of shallow water wave equations, column height and height-averaged horizontal velocity satisfy the same wave equation in the linearized limit, the latter with amplitude
\begin{eqnarray}
u=\sqrt{\frac{g}{d}}h.
\label{EQN_uh}
\end{eqnarray}
For harmonic perturbations, internal dissipation is equivalently described by damping proportional to $u$, and hence by $h$ as a consequence of (\ref{EQN_uh}). 
In Table 2, this is included in the scaling by tidal amplitude. In scaling damping of ocean 
waves by $hF$, the same would be included approximately to within 30\%.

{In considering (\ref{EQN_r4e}) as an effective description of the dominant dissipation in semidiurnal tidal interactions, higher order tidal modes \citep{doo21} are neglected with generally complex dependence on ocean basin geometry \citep{mun66}. In a linearized approximation, one might contemplate including damping by higher harmonics.
However, the latter are difficult to constrain observationally based on data of the Moon's migration velocity. Even if known, evolution of ocean basins on the geological time scale of $10^2$ Myr \citep[e.g.][]{wil66,wil75} makes estimation of time-averages over a recent epoch highly uncertain. For this reason and the expectation that dissipation in the semi-diurnal tides are dominant, these higher order perturbations fall outside the scope of the present approach.}

Numerical integration of (\ref{EQN_r4e}) with (\ref{EQN_epsD}) reproduces a Moon migration time equal to its age for a present $Q$-factor (\ref{EQN_QER}) in excellent agreement with the observational constraint (\ref{EQN_QE}). By (\ref{EQN_epsD}), it introduces a dynamic $Q$-factor
\begin{eqnarray}
Q= \frac{1}{2\min\left\{1, \epsilon_0 hF \right\},}
\end{eqnarray}
that was substantially below $Q_0=1/(2\epsilon_0)$ in the past.
Its trajectory features a startingly fast eviction of the Moon at or close to the initial, unstable synchronous orbit at $j_0$. A protolunar disk with the same composition as the Earth \citep[e.g.][]{bar14} might require even higher Earth spin rates prior, to facilitate its ejection by a giant impact \citep{can01,cuk12}. If so, the Moon is even more pertinent as a deposit of the Earth's initial spin angular momentum.

{Some uncertainty in the formulation of our model arises from the fact that the ocean eigenfrequency $\omega_0$ (closest to the semidiurnal tidal frequency) is that of the Atlantic ocean. By continental drift, it evolves on aforementioned geological time scale that, in particular, might include intermittent closure \citep{wil66,wil75}. In our numerical formulation, the present state then represents a time-average of a recent epoch, small relative to the Moon's age of 4.5 Gyr, probably satisfying
\begin{eqnarray}
\left<\frac{\omega^\prime}{\omega_0}\right> > \frac{\omega^\prime_0}{\omega_0},
\label{EQN_ave}
\end{eqnarray}
where the right hand side refers to today's instantaneous value. Here, we assume that during closure, $\omega_0$ is determined by the remaining oceans, also since $Q$ of the Atlantic near closure is probably suppressed by inversion \citep[uplift of the ocean basin;][]{wil75}). In Fig. \ref{figKE}, (\ref{EQN_ave}) corresponds to moving along the ordinate to the right. By the relative flatness of the curves shown, our main conclusions hold, provided (\ref{EQN_ave}) remains in the window 
\begin{eqnarray}
0.95<\left< \frac{\omega^\prime}{\omega_0}\right> <1.35
\label{EQN_ave2}
\end{eqnarray} 
for solutions to exist. Implications of larger perturbations fall outside the scope of the present formulation.}
	
Once oceans form, eviction of the Moon was swift with steep spindown of the Earth from what was very similar to that of Jupiter today.  In contrast to Jupiter's extreme weather patterns \citep{som88}, our dramatically reduced spin with small Coriolis forces today facilitates a clement global climate. It is tempting to consider application of (\ref{EQN_epsD}) further to exoplanet-moon systems \citep{kip09,for11,sch11,ea16}, whose selection is currently based on temperature, mass and size \citep{men06}. Relatively distant moons may be indirect evidence for effective tidal interaction with oceans, further favoring a potentially clement climate conducive to advanced life.

%{\bf Acknowledgments.} The author gratefully thanks Dr. Bruce Bills for detailed and constructive comments which greatly improved the present manuscript and for pointing out several pertinent references. This report was supported in part by the National Research Foundation of Korea under grant No. 2015R1D1A1A01059793 and 2016R1A5A1013277.
{\bf Acknowledgments.} The author gratefully thanks the reviewer for constructive comments and pointing out several pertinent references. This report was supported in part by the National Research Foundation of Korea under grant No. 2015R1D1A1A01059793 and 2016R1A5A1013277.

%\clearpage

% Don't change these lines
\bsp	% typesetting comment
\label{lastpage}
\end{document}